# APRAP: Another Privacy Preserving RFID Authentication Protocol

Atsuko Miyaji
School of Information Science
Japan Advanced Institute of Science and Technology
1-1 Asahidai, Nomi, Ishikawa, Japan
Email: miyaji@jaist.ac.jp

Mohammad Shahriar Rahman
School of Information Science
Japan Advanced Institute of Science and Technology
1-1 Asahidai, Nomi, Ishikawa, Japan
Email: mohammad@jaist.ac.jp

***Abstract*—Privacy preserving RFID (Radio Frequency Identification) authentication has been an active research area in recent years. Both forward security and backward security are required to maintain the privacy of a tag, i.e., exposure of a tag's secret key should not reveal the past or future secret keys of the tag. We envisage the need for a formal model for backward security for RFID protocol designs in shared key settings, since the RFID tags are too resource-constrained to support public key settings. However, there has not been much research on backward security for shared key environment since Serge Vaudenay in his Asiacrypt 2007 paper showed that perfect backward security is impossible to achieve without public key settings. We propose a Privacy Preserving RFID Authentication Protocol for shared key environment, APRAP [1], which minimizes the damage caused by secret key exposure using insulated keys. Even if a tag's secret key is exposed during an authentication session, forward security and 'restricted' backward security of the tag are preserved under our assumptions. The notion of 'restricted' backward security is that the adversary misses the protocol transcripts which are needed to update the compromised secret key. Although our definition does not capture perfect backward security, it is still suitable for effective implementation as the tags are highly mobile in practice. We also provide a formal security model of APRAP. Our scheme is more efficient than previous proposals from the viewpoint of computational requirements.**

## I. Introduction

One of the main issues of RFID security and privacy has to do with malicious tracking of RFID-equipped objects. While tracking RFID tags is typically one of the key features and goals of a legitimate RFID system, unauthorized tracking of RFID tags is viewed as a major privacy threat. Both forward and backward security are required to maintain the privacy of the tag. Forward security means that even if the adversary acquires the secret data stored in a tag, the tag cannot be traced back using previously known messages [1], [8]. Backward security means the opposite, i.e., even if the adversary acquires the secret data stored in a tag, the tag cannot be traced using subsequently known messages. In other words, exposure of a tag's secret should not reveal any secret information regarding the past or the future of the tag. Moreover, indistinguishability means that the values emitted by one tag should not be distinguishable from the values emitted by other tags [8], [10].

[1]This study is partly supported by Grant-in-Aid for Scientific Research (C), 20500075.

### A. Related Work

Many privacy-preserving mutual RFID authentication schemes have been proposed in recent years [4], [5], [6], [9], [16], [14]. An authentication protocol for RFID from EPCGlobal Class-1 Gen-2 standards was introduced by [5]. Both the authentication key and the access key are updated after a successful session in order to provide forward security. However, [16] showed that [5] is not backward- and forward-secure, because an attacker that compromises a tag can identify a tag's past interactions from the previous communications and the fixed EPC of the tag, and can also read the tag's future transactions. There are also some other privacy-preserving RFID protocols that address untraceability and forward security [4], [6], [14]. However, all these protocols have the same drawback, that is, they cannot provide backward security. LK and SM schemes [9], [16] have recently described RFID authentication schemes satisfying both forward and backward security. However, [16] has been shown to be vulnerable to an attack where an adversary breaks the forward security [15]. The scheme proposed in [9] cannot provide backward security if the current secret key is compromised [11]. Since the adversary is able to trace the target tag at least during the authentication immediately following compromise of the tag secret, perfect backward security makes no sense. Therefore, a minimum restriction should be imposed to achieve backward security, such that the adversary misses the necessary protocol transcripts to update the compromised key. Although this assumption for backward security is true for certain classes of privacy-preserving RFID protocols (i.e., for shared key environment), it is clearly not true for some other cases. For instance, Vaudenay shows an RFID protocol based on public-key cryptography that is resistant to this attack [18]. However, our notion of backward security is true for privacy-preserving RFID protocols based on shared secrets that are updated on each interaction between tag and reader, which is the focus of this paper. Backward security is thus harder to achieve than forward security in general, particularly under the very constrained environment of RFID tags. However, backward security is never less important than forward security in RFID systems. In the case of target tracing, it suffices to somehow steal the tag secret of a target and collect interaction messages to trace the future behaviors of the particular target. Without backward security, this kind of target tracing is trivial. In the case of supply chain management systems, even a catastrophic scenario may take place without backward security: if tag secrets are leaked at some point of tag deployment or during their time in the environment, then all such tags can be traced afterwards. We thus envisage the need for a formal model for backward security in RFID protocol designs (even if not perfect), in addition to the well-recognized forward security.

### B. Our Contribution

We propose APRAP, a privacy-preserving mutual RFID authentication protocol for shared key environment which provides both forward and 'restricted' backward security through key insulation. Even if a tag's secret key is exposed during an authentication session, forward security and 'restricted' backward security of the tag are preserved under our assumptions. The notion of 'restricted' backward security is that the adversary misses the protocol transcripts needed to update the compromised secret key. The protocol also provides indistinguishability between the responses of tags in order to provide privacy of a tag. We also provide a formal security model to design our privacy-preserving protocol. Our assumptions for indistinguishability, and forward/restricted backward security are similar to the assumptions made in previous work.

**Organization of the paper:** The remainder of this paper is organized as follows: Section II presents the notations, assumptions, the protocol model, and the security definitions. Section III describes the protocol. Next, our scheme is evaluated in Section IV based on a security analysis and a comparison with previous work. Section V includes concluding remarks.

## II. Preliminary

### A. Notations

We use the following notations in the protocol description. $H$- a one-way hash function, such that $H : \{0,1\}^* \rightarrow \{0,1\}^\lambda$. $x_i^t$ and $x_i^s$ are $\lambda$-bit random numbers generated during time period $i$ by a tag and a server, respectively. $x^{rand}$ is a $\lambda$-bit random number generated by a server. $sk_i$ is a $\lambda$-bit session key between a tag and a server during time period $i$. $k_i$ is a $\lambda$-bit random shared secret key between a tag and a server during time period $i$. $SK^*$ is a tag-specific master secret key, stored by a legitimate server only. $x_i$ is $\lambda$-bit, generated from $SK^*$ by the server during session $i$. $\oplus$ and $||$ are bitwise XOR operation and concatenation of two bit strings, respectively. $\wr\wr$ represents dividing a bit string into two equal parts.

### B. Assumptions

A tag $\mathcal{T}$ is not tamper-resistant. Initially, it stores the secret key $k_1$ which is updated after each authentication session. All communication between a server and a reader is assumed to be over a private and authentic channel. In this paper, we consider Reader and Server as a single entity. Therefore, we use the terms 'Server' or 'S' interchangeably in the text. The adversary cannot compromise the server. The tag is assumed to be vulnerable to repeated key exposures; specifically, we assume that up to $t < N$ periods can be compromised. Our goal is to minimize the effect such compromises will have. When a secret key is exposed, an adversary will be able to trace the tag for period $i$ until the next single secure authentication session. Our notion of security is that this is the best an adversary can do. In particular, the adversary will be unable to trace a tag for any of the subsequent periods. It is assumed that hash and PRNG take the same amount of execution time. Splitting and concatenation operations take negligible amounts of time.

### C. The Model

We design the model following the model proposed in [7]. However, our model is slightly different than that in [7]. We assume a fixed, polynomial-size tag set $\mathcal{TS} = \{\mathcal{T}^1, \cdots, \mathcal{T}^n\}$, and a server 'Server' as the elements of an RFID system. A Server has information for $\mathcal{TS}$'s authentication such as tag's secret key, master key, etc. Before the protocol is run for the first time, an initialization phase occurs in both $\mathcal{T}^l$ and Server, where $l = 1, \cdots, n$. That is, each $\mathcal{T}^l \in \mathcal{TS}$ runs an algorithm $\mathcal{G}$ to generate the secret key $k^l$, and Server also saves these values in a database field. A key-updating authentication scheme is a 5-tuple of poly-time algorithms ($\mathcal{G}$,$\mathcal{U}^*$,$\mathcal{S}$,$\mathcal{U}$, Auth(AuthT/AuthS) such that:

$\mathcal{G}$, the key generation algorithm, is a probabilistic algorithm which takes as input a security parameter $1^\lambda$, and the total number of tags $n$. It returns a master key $SK^*$, and an initial shared key $k_1$ for each tag.

$\mathcal{U}^*$, the partial key generation algorithm, is a deterministic algorithm which takes as input an index $i$ for a time period (throughout, we assume $1 \leq i \leq N$), the master key $SK^*$ and the secret key $k_i$ of a tag. It returns the partial secret key $x_i$, for time period $i$.

$\mathcal{S}$, the session key generation algorithm, is a deterministic algorithm which takes as input an index $i$, part of the tag's secret key $k_i'$, and a part of the partial secret key $x_i'$. It returns a shared session secret key $sk_i$ for time period $i$.

$\mathcal{U}$, the tag key-update algorithm, is a deterministic algorithm which takes as input an index $i$, part of the tag's secret key $k_i''$, a part of the partial secret key $x_i''$, and a random $x_i^s$. It returns the tag's secret key $k_{i+1}$ for time period $i+1$ (and erases $k_i$, $x_i$, $x_i^s$).

Auth(AuthT/AuthS), the authentication message verification algorithm, is a deterministic algorithm for a server (resp. tag) which takes as input AuthT (resp. AuthS). It returns 1 or the special symbol $\perp$. AuthT/AuthS is as follows:

- AuthT/AuthS, the Tag (resp. Server) authentication message generation algorithm, is a probabilistic algorithm for a tag (resp. server) which takes as input a shared secret $sk_i$, a time period $i$, and random numbers $x_i^t$ and $x_i^s$(or $x^{rand}$) ($k_i', x_i, x_i^s$ (or $x^{rand}$), and $x_i^t$ are the inputs for the server). It returns $\sigma_i'$ (resp. $\sigma_i$) .

APRAP is used as one might expect. A server begins by generating $(SK^*, k_1) \leftarrow \mathcal{G}(1^\lambda, n)$, storing $SK^*$ on a server (physically-secure device), and storing $k_1$ in both the server and the tag. At the beginning of time period $i$, the tag requests $x_i = \mathcal{U}^*(i, SK^*, k_i)$ from the server. Using $x_i$, and $k_i$, the tag may compute the session secret key $sk_i = \mathcal{S}(i, k'_i, x'_i)$. This key is used to create authentication messages sent during time period $i$. Both the tag and server update their shared secret by $k_{i+1} = \mathcal{U}(i, k''_i, x''_i, x^s_i)$. After computation of $k_{i+1}$, the tag must erase $k_i$, and $x_i$.

### D. Security Definitions

Adversary $\mathcal{A}$'s interaction with the RFID entities in the network is modeled by sending the following queries to an oracle $\mathcal{O}$ and receiving the result from $\mathcal{O}$. The queries in our model follow [8] with some differences. We do not need Reply*/Execute*, since we do not consider a tag to be maintaining an internal state in our protocol. Also, we consider server and reader as a single entity. So, we do not need Forward$_1$/Forward$_2$ and Auth queries. Instead, Reply, Reply' perform the tasks of Forward$_1$, Forward$_2$, respectively. They also serve the purpose of Auth(AuthT/AuthS).

- *Query*$(S, x^s_i)$: It calls server ($S$) and outputs $x^s_i$ of period $i$.
- *Query'*$(\mathcal{T}^l_i, x^t_i)$: It calls tag ($\mathcal{T}^l$) and outputs $x^t_i$ of period $i$.
- *Query$_b$*$(S, x^{rand})$: It calls server ($S$) and outputs any random $x^{rand}$.
- *Reply*$(S, x^t_i, \sigma_i, \delta_i)$: It calls $S$ with input $x^t_i$ and outputs $\sigma_i, \delta_i$ for period $i$. It uses AuthS algorithm. The output is forwarded to $\mathcal{T}^l$.
- *Reply'*$(\mathcal{T}^l_i, x^s_i, \sigma_i, \delta_i, \sigma'_i)$: It calls $\mathcal{T}^l$ with input $x^s_i, \sigma_i, \delta_i$ and outputs $\sigma'_i$ for period $i$. It uses AuthT algorithm. The output is forwarded to $S$.
- *Reply$_b$*$(\mathcal{T}^l_i, x^{rand}, \sigma_i, \delta_i, \sigma'_i)$: It calls $\mathcal{T}^l$ with input $x^{rand}, \sigma_i, \delta_i$ and outputs $\sigma'_i$ for period $i$. It uses AuthT algorithm. The output is forwarded to $S$.
- *Execute*$(\mathcal{T}^l_i, S)$: This query uses the algorithms ($\mathcal{G}, \mathcal{U}^*, \mathcal{S}, \mathcal{U}$, Auth(AuthT/AuthS)). It receives the protocol transcripts $\sigma_i, x^s_i, \sigma'_i, \delta_i, x^t_i$, and outputs them. This models the adversary $\mathcal{A}$'s eavesdropping of protocol transcripts. It has the following relationships with the above queries: *Execute* $(\mathcal{T}^l_i, S)$ = *Query*$(S, x^s_i)$ $\wedge$ *Query'*$(\mathcal{T}^l_i, x^t_i)$ $\wedge$ *Reply*$(S, x^t_i, \sigma_i, \delta_i)$ $\wedge$ *Reply'*$(\mathcal{T}^l_i, x^s_i, \sigma_i, \delta_i, \sigma'_i)$.
- *Execute$_b$*$(\mathcal{T}^l_i, S)$: This query uses the algorithms ($\mathcal{G}, \mathcal{U}^*, \mathcal{S}, \mathcal{U}$, Auth(AuthT/AuthS)). It receives the protocol transcripts $\sigma_i, \sigma'_i, \delta_i, x^t_i, x^{rand}$, and outputs them. This models the adversary $\mathcal{A}$'s eavesdropping of protocol transcripts except $x^s_i$ which is used for key update. It has the following relationship with the above queries: *Execute$_b$*$(\mathcal{T}^l_i, S)$ = *Query$_b$*$(\mathcal{T}^l_i, x^{rand})$ $\wedge$ *Query'*$(\mathcal{T}^l_i, x^t_i)$ $\wedge$ *Reply*$(S, x^t_i, \sigma_i, \delta_i)$ $\wedge$ *Reply$_b$*$(\mathcal{T}^l_i, x^{rand}, \sigma_i, \delta_i, \sigma'_i)$.
- *RevealSecret* $(\mathcal{T}^l, i)$: This query uses the algorithm $\mathcal{U}$. It receives the tag's $\mathcal{T}^l$ secret key $k_i$, and outputs $k_i$ of period $i$.
- *Test* $(\mathcal{T}^l, i)$: This query is allowed only once, at any time during $\mathcal{A}$'s execution. A random bit $b$ is generated; if $b = 1$, $\mathcal{A}$ is given transcripts corresponding to the tag, and if $b = 0$, $\mathcal{A}$ receives a random value.

We now give the definitions through security games, reminiscent of classic indistinguishability in a cryptosystem security game. We follow [8] to define indistinguishability and forward security. The success of $\mathcal{A}$ in the games is subject to $\mathcal{A}$'s advantage in distinguishing whether $\mathcal{A}$ has received an RFID tag's real response or a random value. The next two games represent the attack games for forward security and restricted backward security, respectively.

*Definition 1:* Indistinguishability

- Phase 1: Initialization

(1) Run algorithm $\mathcal{G}(1^\lambda, n) \to (k^1, \ldots, k^n)$.
(2) Set each tag $\mathcal{T}^l$'s secret key as $k^l$, where $\mathcal{T}^l \in \mathcal{TS} = \{\mathcal{T}^1, \ldots, \mathcal{T}^n\}$.
(3) Save each $\mathcal{T}^l$'s $k^l$ generated in step (1) in Server's field.

- Phase 2: Learning

(1) $\mathcal{A}^{ind}$ executes *Query*$(S, x^s_i)$, *Query'*$(\mathcal{T}^l_i, x^t_i)$, *Reply*$(S, x^t_i, \sigma_i, \delta_i)$, *Reply'*$(\mathcal{T}^l_i, x^s_i, \sigma_i, \delta_i, \sigma'_i)$, and *Execute* $(\mathcal{T}^l_i, S)$ oracles for all $n-1$ tags, except the $\mathcal{T}^c \in \mathcal{TS}$ used in challenge phase.

- Phase 3: Challenge

(1) $\mathcal{A}^{ind}$ selects a challenge tag $\mathcal{T}^c$ from $\mathcal{TS}$.
(2) $\mathcal{A}^{ind}$ executes *Query*$(S, x^s_i)$, *Query'*$(\mathcal{T}^l_i, x^t_i)$, *Reply*$(S, x^t_i, \sigma_i, \delta_i)$, *Reply'*$(\mathcal{T}^l_i, x^s_i, \sigma_i, \delta_i, \sigma'_i)$, and *Execute* $(\mathcal{T}^l_i, S)$ oracles for $\mathcal{T}^c$, where $i = 1, \ldots, q-1$.
(3) $\mathcal{A}^{ind}$ calls the oracle *Test*$(\mathcal{T}^c, i)$.
(4) For the $\mathcal{A}^{ind}$'s *Test*, Oracle $\mathcal{O}$ tosses a fair coin $b \in \{0, 1\}$; let $b \xleftarrow{R} \{0, 1\}$.
i. If $b = 1$, $\mathcal{A}^{ind}$ is given the messages corresponding to $\mathcal{T}^c$'s $i$-th instance.
ii. If $b = 0$, $\mathcal{A}^{ind}$ is given random values.
(5) $\mathcal{A}^{ind}$ outputs a guess bit $b'$.

$\mathcal{A}$ wins if $b = b'$

The advantage of any PPT adversary $\mathcal{A}^{ind}$ with computational boundary $e_1, r_1, r_2, \lambda$, where $e_1$ is the number of *Execute*, $r_1$ is the number of *Reply*, $r_2$ is the number of *Reply'* and $\lambda$ is the security parameter, is defined as follows:

$$Adv^{ind}_{\mathcal{A}^{ind}} = |Pr[b = b'] - 1/2|$$

The scheme provides indistinguishability if and only if the advantage of $Adv^{ind}_{\mathcal{A}^{ind}}$ is negligible.

*Definition 2:* Forward Security

- Phase 1: Initialization

(1) Run algorithm $\mathcal{G}(1^\lambda, n) \to (k^1, \ldots, k^n)$.
(2) Set each tag $\mathcal{T}^l$'s secret key as $k^l$, where $\mathcal{T}^l \in \mathcal{TS} = \{\mathcal{T}^1, \ldots, \mathcal{T}^n\}$.
(3) Save each $\mathcal{T}^l$'s $k^l$ generated in step (1) in Server's field.

- Phase 2: Learning

(1) $\mathcal{A}^{for}$ executes *Query*$(S, x^s_i)$, *Query'*$(\mathcal{T}^l_i, x^t_i)$, *Reply*$(S, x^t_i, \sigma_i, \delta_i)$, *Reply'*$(\mathcal{T}^l_i, x^s_i, \sigma_i, \delta_i, \sigma'_i)$, and *Execute* $(\mathcal{T}^l_i, S)$ oracles for all $n-1$ tags, except for the $\mathcal{T}^c \in \mathcal{TS}$ used in challenge phase.

- Phase 3: Challenge

(1) $\mathcal{A}^{for}$ selects a challenge tag $\mathcal{T}^c$ from $\mathcal{TS}$.
(2) $\mathcal{A}^{for}$ executes *Query*$(S, x^s_i)$, *Query'*$(\mathcal{T}^l_i, x^t_i)$, *Reply*$(S, x^t_i, \sigma_i, \delta_i)$, *Reply'*$(\mathcal{T}^l_i, x^s_i, \sigma_i, \delta_i, \sigma'_i)$, *Execute* $(\mathcal{T}^l_i, S)$, and *RevealSecret*$(\mathcal{T}^c, i)$ oracles for $\mathcal{T}^c$ for $\mathcal{T}^c$'s $i$-th instance.
(3) $\mathcal{A}^{for}$ calls the oracle *Test*$(\mathcal{T}^c, i-1)$.
(4) For the $\mathcal{A}^{for}$'s *Test*, Oracle $\mathcal{O}$ tosses a fair coin $b \in \{0, 1\}$; let $b \xleftarrow{R} \{0, 1\}$.

i. If $b = 1$, $\mathcal{A}^{for}$ is given the messages corresponding to $\mathcal{T}^c$'s $(i-1)$-th instance.
ii. If $b = 0$, $\mathcal{A}^{for}$ is given random values.
(5) $\mathcal{A}^{for}$ executes the oracles for $n-1$ tags of $\mathcal{TS}$, except $\mathcal{T}^c$, like in the learning phase.
(6) $\mathcal{A}^{for}$ outputs a guess bit $b'$.

$\mathcal{A}$ wins if $b = b'$

The advantage of any PPT adversary $\mathcal{A}^{for}$ with computational boundary $e_1, r_1, r_2, \lambda$, where $e_1$ is the number of *Execute*, $r_1$ is the number of *Reply*, $r_2$ is the number of *Reply'* and $\lambda$ is the security parameter, is defined as follows:

$$Adv^{for}_{\mathcal{A}^{for}} = |Pr[b = b'] - 1/2|$$

The scheme is forward secure if and only if the advantage of $Adv^{for}_{\mathcal{A}^{for}}$ is negligible.

*Definition 3:* Restricted Backward Security [1]

- Phase 1: Initialization

(1) Run algorithm $\mathcal{G}(1^\lambda, n) \rightarrow (k^1, \ldots, k^n)$.
(2) Set each tag $\mathcal{T}^l$'s secret key as $k^l$, where $\mathcal{T}^l \in \mathcal{TS} = \{\mathcal{T}^1, \ldots, \mathcal{T}^n\}$.
(3) Save each $\mathcal{T}^l$'s $k^l$ generated in step (1) in Server's field.

- Phase 2: Learning

(1) $\mathcal{A}^{back}$ executes *Query*$_b$($\mathcal{T}^l_i, x^{rand}$), *Query'*($\mathcal{T}^l_i, x^t_i$), *Reply*($S, x^t_i, \sigma_i, \delta_i$), *Reply*$_b$($\mathcal{T}^l_i, x^{rand}, \sigma_i, \delta_i, \sigma'_i$), and *Execute*$_b$($\mathcal{T}^l_i, S$) oracles for all $n-1$ tags, except for the $\mathcal{T}^c \in \mathcal{TS}$ used in challenge phase.

- Phase 3: Challenge

(1) $\mathcal{A}^{back}$ selects a challenge tag $\mathcal{T}^c$ from $\mathcal{TS}$.
(2) $\mathcal{A}^{back}$ executes *Query*$_b$($\mathcal{T}^l_i, x^{rand}$), *Query'*($\mathcal{T}^l_i, x^t_i$), *Reply*($S, x^t_i, \sigma_i, \delta_i$), *Reply*$_b$($\mathcal{T}^l_i, x^{rand}, \sigma_i, \delta_i, \sigma'_i$), *Execute*$_b$($\mathcal{T}^l_i, S$), and *RevealSecret*($\mathcal{T}^c, i$) oracles for $\mathcal{T}^c$'s $i$-th instance.
(3) $\mathcal{A}^{back}$ calls the oracle *Test*($\mathcal{T}^c, i+1$).
(4) For the $\mathcal{A}^{back}$'s *Test*, Oracle $\mathcal{O}$ tosses a fair coin $b \in \{0, 1\}$; let $b \xleftarrow{R} \{0, 1\}$.
i. If $b = 1$, $\mathcal{A}^{back}$ is given the messages corresponding to $\mathcal{T}^c$'s $i+1$th instance.
ii. If $b = 0$, $\mathcal{A}^{back}$ is given random values.
(5) $\mathcal{A}^{back}$ executes oracles for $n-1$ tags of $\mathcal{TS}$, except $\mathcal{T}^c$, like in the learning phase.
(6) $\mathcal{A}^{back}$ outputs a guess bit $b'$.

$\mathcal{A}$ wins if $b = b'$

The advantage of any PPT adversary $\mathcal{A}^{back}$ with computational boundary $e_2, r_1, r_b, \lambda$, where $e_2$ is the number of *Execute*$_b$, $r_1$ is the number of *Reply*, $r_b$ is the number of *Reply*$_b$ and $\lambda$ is the security parameter, is defined as follows:

$$Adv^{for}_{\mathcal{A}^{back}} = |Pr[b = b'] - 1/2|$$

[1]Since once obtaining the tag secret by *RevealSecret*, $\mathcal{A}^{back}$ takes all the power of the tag itself and thus can trace the target tag at least during the authentication immediately following the attack. In typical RFID system environments, tags and readers operate only at short communication range and for a relatively short period of time. Thus, the minimum restriction for backward security is such that the adversary misses the protocol transcripts needed to update the compromised secret key. The same restriction was applied in [16]. On the other hand, [9] claimed that there should exist some non-empty gap not accessible by the adversary between the time of a reveal query and the attack time. But this restriction was shown to be inadequate to provide backward security by [11].

The scheme is restricted backward secure if and only if the advantage of $Adv^{back}_{\mathcal{A}^{back}}$ is negligible.

*Definition 4:* Privacy-Preserving Protocol

A protocol is privacy-preserving when indistinguishability, forward security, and restricted backward security are guaranteed for any PPT adversary $\mathcal{A}$ with computational boundary $e_1, r_1, e_2, r_2, r_b, \lambda$, where $e_1$ is the number of *Execute*, $r_1$ is the number of *Reply*, $e_2$ is the number of *Execute*$_b$, $r_2$ is the number of *Reply'*, $r_b$ is the number of *Reply*$_b$ and $\lambda$ is the security parameter.

## III. Protocol Description

Table I describes the protocol building blocks, and Fig. 1 describes the authentication session. During any session $i$, the following steps take place between a tag and a server:

1. The server sends a random challenge $x^s_i$ to the tag.
2. The tag replies to the server with a random $x^t_i$.
3. The server splits $k_i$ into $k'_i$ and $k''_i$, and $x^s_i$ into $x^{s'}_i$ and $x^{s'}_i$. It then generates $x_i$ from $SK^*$ and $k_i$ by $H_i(SK^*, k_i)$, where $H_i$ is the $i$-th time run for $H$. $SK^*$ is used to generate $x_i$ so that no other entities other than a valid server can generate $x_i$. Even if an adversary compromises $k_i$, it can not generate $x$ for any subsequent sessions using only that $k_i$. $x^s_i$ is used as a random number for server authentication, and $x_i$ is used as the partial key for the present session. The server computes $\sigma_i = H(k'_i \| x_i, x^s_i \| x^t_i)$, and $\delta_i = k_i \oplus x_i$. The server sends $\sigma_i$ and $\delta_i$ to the tag.
4. After receiving $\sigma_i$ and $\delta_i$, the tag splits $k_i$ into $k'_i$ and $k''_i$, and extracts $x_i$ from $\delta_i$. The tag then authenticates the server by verifying $\sigma_i$. If the server is authenticated as a legitimate server, the tag splits $x^s_i$ into $x^{s'}_i$ and $x^{s''}_i$, and $x_i$ into $x'_i$ and $x''_i$. The tag now computes the session secret key $sk_i$ by concatenating $k'_i$ and $x'_i$. It then computes $\sigma'_i = H(x^t_i \| x^s_i, sk_i)$, and updates its own secret key to $k_{i+1}$ by $H(k''_i \| x''_i, x^s_i)$. The tag sends $\sigma'_i$ to the server, and erases $x_i$, $x^t_i$, and $sk_i$ from its memory. The updated $k_{i+1}$ is used for the next authentication session.
5. After the server receives $\sigma'_i$, it authenticates the tag by verifying $\sigma'_i$. The server then updates the secret key to $k_{i+1}$ of the tag by $H(k''_i \| x''_i, x^s_i)$. This updated $k_{i+1}$ is stored in the server database, and is used for the next authentication session.[2]

## IV. Evaluation

### A. Security Analysis

Due to page limitation, we omit the security proofs and put them in the full version.

*Theorem 1:* The protocol $\pi$ = ($\mathcal{G}$, $\mathcal{U}^*$, $\mathcal{S}$, $\mathcal{U}$, Auth(AuthT/AuthS) provides indistinguishability for any PPT adversary $\mathcal{A}^{ind}$ with computational boundary $e_1, r_1, r_2, \lambda$, where $e_1$ is the number of *Execute*, $r_1$ is the number of *Reply*, $r_2$ is the number of *Reply'* and $\lambda$ is the security parameter.

[2]Note that it is imperative for the respective times taken by authentication success and failure to be as close as possible to prevent obvious timing attacks by malicious readers (aimed at distinguishing among the two cases)[17]. For this reason, even if the authentication by a tag is failed, it should generate random numbers instead of simply failure, to make the cases of success and failure indistinguishable from each other.

TABLE I
PROTOCOL BUILDING BLOCKS

| | |
|---|---|
| $\mathcal{U}^*$: | Auth (AuthT/ AuthS) |
| input: $i, SK^*, k_i$ | AuthT: |
| compute: $H_i(SK^*, k_i)$ | input: $i, x_i^t, x_i^s, sk_i$ |
| return: $x_i$ | compute: $H(x_i^t \| x_i^s, sk_i)$ |
| $\mathcal{S}$: | return: $\sigma'_i$ |
| input: $i, k'_i, x'_i$ | AuthS: |
| compute: $k'_i \| x'_i$ | input: $i, x_i^s, k'_i, x_i, x_i^t$ |
| return: $sk_i$ | compute: $H(k'_i \| x_i, x_i^s \| x_i^t)$ |
| $\mathcal{U}$: | return: $\sigma_i$ |
| input: $i, k''_i, x''_i, x_i^s)$ | return: 1 or $\perp$ |
| compute: $H(k''_i \| x''_i, x_i^s)$ | |
| return: $k_{i+1}$ | |

| **Tag:** | | **Server:** |
|---|---|---|
| $k_i$ | | $SK^*, k_i$ |
| | | $x_i^s \in \{0,1\}^*$ |
| | $\xleftarrow{x_i^s}$ | |
| $x_i^t \in \{0,1\}^*$ | | |
| | $\xrightarrow{x_i^t}$ | |
| | | $k_i = k'_i \wr\wr k''_i$ |
| | | $x_i^s = x_i^{s'} \wr\wr x_i^{s''}$ |
| | | $\mathcal{U}^*(i, SK^*, k_i) \rightarrow x_i$ |
| | | AuthS$(i, k'_i, x_i, x_i^s, x_i^t) \rightarrow \sigma_i$ |
| | | $\delta_i = k_i \oplus x_i$ |
| | $\xleftarrow{\sigma_i, \delta_i}$ | |
| $k_i = k'_i \wr\wr k''_i$ | | |
| $x_i = \delta_i \oplus k_i$ | | |
| Auth(AuthS)$\rightarrow$ 1 or $\perp$ | | |
| $x_i^s = x_i^{s'} \wr\wr x_i^{s''}$ | | |
| $x_i = x'_i \wr\wr x''_i$ | | |
| $\mathcal{S}(i, k'_i, x'_i) \rightarrow sk_i$ | | |
| AuthT$(i, x_i^t, x_i^s, sk_i) \rightarrow \sigma'_i$ | | |
| $\mathcal{U}(i, k''_i, x''_i, x_i^s) \rightarrow k_{i+1}$ | | |
| | $\xrightarrow{\sigma'_i}$ | |
| | | Auth(AuthT)$\rightarrow$ 1 or $\perp$ |
| | | $\mathcal{U}(i, k''_i, x''_i, x_i^s) \rightarrow k_{i+1}$ |

Fig. 1. Our Scheme: APRAP

*Theorem 2:* The protocol $\pi$ = ($\mathcal{G}$, $\mathcal{U}^*$, $\mathcal{S}$, $\mathcal{U}$, Auth(AuthT/ AuthS) is forward secure for any PPT adversary $\mathcal{A}^{for}$ with computational boundary $e_1, r_1, r_2, \lambda$, where $e_1$ is the number of *Execute*, $r_1$ is the number of *Reply*, $r_2$ is the number of *Reply'* and $\lambda$ is the security parameter.

*Theorem 3:* The protocol $\pi$ = ($\mathcal{G}$, $\mathcal{U}^*$, $\mathcal{S}$, $\mathcal{U}$, Auth(AuthT/ AuthS) is restricted backward secure for any PPT adversary $\mathcal{A}^{back}$ with computational boundary $e_2, r_1, r_b, \lambda$, where $e_2$ is the number of *Execute$_b$*, $r_1$ is the number of *Reply*, $r_b$ is the number of *Reply$_b$* and $\lambda$ is the security parameter.

*Theorem 4:* The protocol $\pi$ = ($\mathcal{G}$, $\mathcal{U}^*$, $\mathcal{S}$, $\mathcal{U}$, Auth(AuthT/ AuthS) is privacy-preserving for any PPT adversary $\mathcal{A}$ with computational boundary $e_1, e_2, r_1, r_2, r_b, \lambda$, where $e_1$ is the number of *Execute*, $e_2$ is the number of *Execute$_b$*, $r_1$ is the number of *Reply*, $r_2$ is the number of *Reply'*, $r_b$ is the number of *Reply$_b$* and $\lambda$ is the security parameter.

### *B. Discussion and Comparison With Previous Work*

Deursen et al. [21] discussed a weakness of the indistinguishability definition of [8]. Deursen et al. argued that, to achieve location privacy, the adversary must not be able to distinguish one tag's response from other tags' responses, but it is not necessary that the adversary cannot distinguish the tag's response from any arbitrary value. However, our definition can be modified according to their argument. For that purpose, the oracle queries should run on all but two tags which are used for the challenge phase. All the adversary needs to do is to distinguish between those two tags. In fact, our assumption about the tag responses is such that the output of the one-way hash functions are indistinguishable from a random bit string of equal length.

In [2], Bellare et al. show that it is impossible to achieve public-channel key insulated security in the face of an active adversary (who can compromise the secret key). Although we follow the idea of key insulation from [7], assuming passive adversary in case of RFID (who can eavesdrop only) is not practical, as it is easy for an adversary to break into a tag's memory. Considering this, the assumptions made in our scheme (as well as in [16]) are more realistic to achieve restricted backward security, and the other features as well. However, many of the existing mutual authentication protocols may support restricted backward security under our assumption ([3], [19], [17] to name a few). But [3], [19] require a tag to remember too many secrets. Moreover, [3], [19] cannot provide forward security as shown by [13] and [22], respectively. Again, [17] requires more computation than our scheme, and it does not provide reader authentication. Nevertheless, none of these protocols came up with a formal model of backward security (even if not perfect).

Although it is not the primary target of our proposed protocol, it is also possible to prevent desynchronization attacks [20] in our protocol to some extent. We consider the following type of attack: If the last message is blocked, the tag updates the shared secret key, $k_i$, but the server doesn't. The server and tag are no longer able to communicate successfully. To prevent such an attack, the server has to remember the last valid authentication session transcripts and the secret values. When a server receives some random number instead of a valid authentication value from a tag, the server updates itself using the information from the last valid session, and tries again to get synchronized with the tag. Although the question of scalability is an issue here, this approach can help avoid such desynchronization attacks in a limited way (of course the system gets desynchronized if the last messages from two consecutive sessions are blocked). Even though the system gets desynchronized, an adversary can not trace a tag from its desynchronized state, since the responses of a tag are always pseudorandom, hence indistinguishable. However, we are more concerned with 'exposure resilience' of the secret key and its effect on the authentication protocol, rather than the desynchronization attacks. Providing full resistance against desynchronization attacks is a separate issue.

We compare our work, based on security properties and computational cost, with LK and SM schemes in Table II below. According to [8], a scheme must satisfy both forward security and indistinguishability in order to achieve 'strong location privacy'. If a scheme satisfies indistinguishability only, the scheme is 'weak location private'. [15] has shown that SM scheme is not forward secure. So, SM scheme is weak location private only, whereas our scheme is strong location private. SM scheme furthermore does not give any formal security model for indistinguishability and forward security. Regarding computational requirements, our protocol requires a simple one-way hash function, random number generation and the XOR operation. We use a simple hash function like SQUASH [12] to achieve forward security for the tag. This requires around 1K gates.

As the server needs to authenticate itself first to a tag, the server must broadcast the authentication messages to the tags. Since the server does not know the id of the tag that it wants to authenticate, the server has to compute and broadcast the authentication messages for all the tags in its storage. We assume that the server has enough resource to perform such computation. On the other hand, a tag receiving the broadcast messages has to find a match with it's verification value. Although computing the verification value is always constant, finding a match increases the required computations according to the number of broadcast messages in the worst case. As stated earlier, such a scenario is unavoidable when we require that a server should authenticate itself first to a tag. We say that our scheme is more suitable for an environment where the reader must read a number of tags at a time (inventory management) and/or where there are not too many tags (library with a few thousand books).

TABLE II
PERFORMANCE COMPARISON WITH PREVIOUS WORKS

| schemes | ind. | for. sec. | back. sec. | tag's comp. | tag's storage |
|---|---|---|---|---|---|
| LK [9] | $\surd$ | $\surd$ | $X$ | 2 XOR, 5 hash | 384 bits |
| SM [16] | $\surd$ | $X$ | $\surd$ | 6 XOR, 4 hash | 128 bits |
| APRAP | $\surd$ | $\surd$ | $\surd$ | 1 XOR, 4 hash | 128 bits |

• assuming each secret key is 128 bits long; hash functions and PRNG require the same computational resources; ind.: indistinguishability; for. sec.: forward security; back. sec.: restricted backward security; $\surd$: the property is satisfied; $X$: the property is not satisfied

## V. CONCLUSION

We have proposed APRAP, a privacy-preserving mutual RFID authentication protocol for shared key environment. The protocol uses two different keys for mutual authentication. The server sends a random partial key (generated from a master secret key $SK^*$) to a tag. The tag generates the session key $sk$ to authenticate itself to the server. The tag's secret key $k$ is updated using a partial key received from the server. As $k$ is purely fresh for every time period, the tag's security is guaranteed for all other time periods (both for the past and future) under our assumptions. We show that our scheme is computationally more efficient than the SM and LK schemes. Our protocol satisfies indistinguishability, and achieves both forward and restricted backward security through key-insulation. We provide a formal security model of the proposed protocol as well.